\documentclass[12pt]{iopart}
\usepackage{graphicx}

\begin{document}

\title{Stray field and superconducting surface spin valve effect  in  La$_{0.7}$Ca$_{0.3}$MnO$_3$/YBa$_2$Cu$_3$O$_{7-\delta}$ bilayers}

\author{T. Hu$^{1}$, H. Xiao$^{1,2}$, C. Visani$^{3}$, J. Santamaria$^{3}$, C. C. Almasan$^{1}$}

\address{$^{1}$ Department of Physics, Kent State University, Kent, Ohio, 44242, USA}

\address{$^{2}$ Beijing National Laboratory for Condensed Matter Physics,
Institute of Physics, Chinese Academy of Sciences, Beijing 100190, China}

\address{$^{3}$ GFMC, Departamento Fisica Aplicada III, Universidad Complutense de Madrid, 28040 Madrid, Spain}

\ead{calmasan@kent.edu}

\begin{abstract}
Electronic transport and magnetization measurements were performed on La$_{0.7}$Ca$_{0.3}$MnO$_3$/YBa$_2$Cu$_3$O$_{7-\delta}$ (LCMO/YBCO) bilayers  below the  superconducting transition temperature in order to study the interaction between magnetism and superconductivity.  This study shows that a substantial number of weakly pinned vortices are  induced  in the YBCO layer by the large out-of-plane stray field in the domain walls. Their motion gives rise to large dissipation peaks at the coercive field. The angular dependent magnetoresistance (MR) data reveal the interaction between  the stripe domain structure present in the LCMO layer and the vortices and anti-vortices induced in the YBCO layer by the out-of-plane stray field.  In addition, this study shows that a superconducting surface spin valve effect is present in these bilayers as a result of the relative orientation between the magnetization at the LCMO/YBCO interface and the magnetization in the interior of the LCMO layer that can be tuned by the rotation of a small $H$. This latter finding will facilitate the development of superconductive magnetoresistive memory devices.  These low-magnetic field MR data, furthermore, suggest that triplet superconductivity is induced in the LCMO layer, which is  consistent  with recent reports of triplet superconductivity in LCMO/YBCO/LCMO trilayers and LCMO/YBCO bilayers. 
\end{abstract}

\maketitle

\section{Introduction}
The physics of spin dependent transport  is currently attracting much attention because of its fundamental interest to the realization of spintronic devices \cite{Prinz}.  A lot of studies were focused on the  ferromagnet/nonmagnetic-spacers/ferromagnet (F/N/F) structures for which the giant magnetoresistance GMR depends on the relative orientation of the magnetization in the top and bottom F layers, giving rise to the  spin valve effect \cite{Baibich} in a GMR memory device.
When the nonmagnetic spacer was replaced with a superconductor, a novel $\it superconducting$ spin valve  effect was proposed and theoretically justified \cite{Tagirov} in ferromagnet/superconductor/ferromagnet (F/S/F) structures, in which the superconductivity is switched on and off by reversing the magnetization direction of one of the ferromagnetic layers. There are several reports of the experimental realization of this effect by using metals as the S layer \cite{Gu, Potenza, Moraru}. 

Recently a  $\it surface$ spin valve effect was observed  within a few atomic layers at the ferromagnetic/nonmagnetic (F/N) interface, which is   due to the fact
that the ferromagnetic spins at such an interface are  significantly different from the magnetic character of the spins inside the F layer and they can act as current- or field-driven spin valves with respect to the magnetization in the interior of the ferromagnetic layer \cite{Yanson}. Based on these results  we anticipate that  a superconducting surface spin valve effect could be present in a ferromagnet/superconductor
 (F/S) bilayer such as La$_{0.7}$Ca$_{0.3}$MnO$_3$/YBa$_2$Cu$_3$O$_{7-\delta}$ (LCMO/YBCO)  because  it has been  shown that the magnetization of the LCMO/YBCO interface is significantly different from the bulk magnetization, inside the LCMO layer  \cite{interface1,interface2,interface3}. This finding will facilitate the development of superconductive magnetoresistive memory devices.

The domain structure of the ferromagnetic layer has  a significant
influence on  the superconductivity of the superconducting layer \cite{Steiner}; i.e.,  both N\'{e}el and Bloch domain walls can enhance or suppress superconductivity  \cite{Rusanov,Rusanov2} depending on the size of the coherence length  of  the Cooper pair ($\xi_{ab}$) relative to the width of the domain wall ($\delta$). In the case of the LCMO/YBCO bilayers, $\xi_{ab} \approx$ 3 nm at 45 K  [with  $\xi_{ab}$(0 K) = 2 nm and the superconducting transition temperature of the LCMO/YBCO bilayer $T_c$ = 82 K] that is much smaller than the width of the domain walls of the LCMO (about 3 $\mu$m and 2 $\mu$m at  63 and 10 K, respectively \cite{interface2}). Therefore, both N\'{e}el and Bloch domain walls suppress superconducitivity in the LCMO/YBCO bilayers due to the effect of the exchange interaction on the Cooper pairs \cite{Buzdin}.  Moreover, the out-of-plane spins in the Bloch domain walls induce vortices, which give rise to additional dissipation \cite{Bell}. 

To study the effect of domain walls on superconductivity and to search for the superconductive surface spin valve effect, we   performed  angular dependent transport measurements on LCMO/YBCO
bilayers by rotating the magnetic field $H$ within the $ab$-plane. This study revealed  that  vortices are induced in these bilayers by the out-of-plane stray field in the domain walls. This latter field is induced by the stresses in the twins of the LCMO layer as a result of a structural phase transition in the substrate. The motion of these vortices gives rise to one type of angular dependent magnetoresistance (MR) dissipation. In addition, the present study shows that one can generate a superconducting surface spin valve effect in these bilayers, in which the MR depends on the relative angle between the magnetizations of the LCMO/YBCO interface and of the LCMO bulk layer. These two types of  behavior were observed in LCMO/YBCO bilayers {\it only} below the superconducting transition temperature $T_c$ of the bilayer and are not present in the normal state. 

\section{Experimental Details}
LCMO/YBCO bilayers were grown on (100)-oriented SrTiO$_3$ single crystals. The details of sample preparation were reported elsewhere \cite{Pena2006}. The ferromagnetic layer of the bilayer  is 40 unit cells (u. c.) (16 nm); the superconducting layer is 4 u. c. (4.8 nm). The  LCMO/YBCO interfaces are sharp and perfectly coherent \cite{Pena2004}.  All samples are $1\times$ 0.5 cm$^2$.  For all the data shown here, a current $I$ of 100 $\mu$A was applied in the $ab$ plane and the resistance $R$ of the bilayer  was measured using a four-contact method. The applied  field $H$ was  rotated in the $ab-$plane and the angle $\varphi$ is defined as the angle between $H$ and  the  [010] direction of the LCMO  layer [see  inset to Fig. 2(a)].  We repeated the measurements with other values of the applied current in the range 1 $\mu$A to 100 $\mu$A and found that the results presented here are qualitatively independent of these values of the applied current.

A small out-of-plane misalignment of $H$  is found when $H$ is rotated in the $ab$ plane. In a one axis rotator system, it is very hard to ensure an in-plane alignment of $H$ of better than about $\pm 3$ degrees. This misalignment, i.e. the magnetic field is not completely within the $ab$ plane of the single crystal, gives an  angular dependent resistance that is  independent of the current direction.  Its magnitude decreases with decreasing field and it has an 180$^0$ periodicity  \cite{xiaoPRB}. All the data shown in this paper are after the subtraction of this misalignment contribution to the resistance. 

\section{Stray field effect} 
\subsection{Magetoresistance and magnetization measurements}
Figure 1 is a plot of the resistance $R$ (open squares) and magnetization $M$ (open circles) of  a LCMO/YBCO bilayer measured at a temperature $T$ of 45 K $< T_c$ [$T_c =82$ K is  defined in the inset (a)
to Fig. 1] vs the magnetic field $H$ applied in the $ab-$plane along the [010] crystallographic direction of the LCMO layer ($\varphi=0$).
The magnetic field is scanned from $-2000$ Oe up to $+2000$ Oe and then  back to $-2000$ Oe.  Two sharp  resistance peaks are
present in the $R(H)$ data measured with $I\parallel$ [100] crystallographic direction. The positions of the resistance peaks are at $+280$ and $-280$ Oe,  during increasing and decreasing  $H$, respectively,  corresponding to the coercive field (zero magnetization)  of the sample,  determined from the $M(H)$ data of this figure. [Notice that, since the magnetization curve is measured in the superconducting state, there is a contribution due to the superconducting moment.] At the coercive field, the LCMO layer has the maximum number of domains, hence domain walls DWs. Therefore, the stay field is the largest. Hence, the fact that the resistance peaks appear exactly at the coercive field of the sample indicates that they are the result of the stay field. Nevertheless the question that needs to be answered next is the direction of the stray field.

\begin{figure}
\centering
\includegraphics[width=0.7\textwidth]{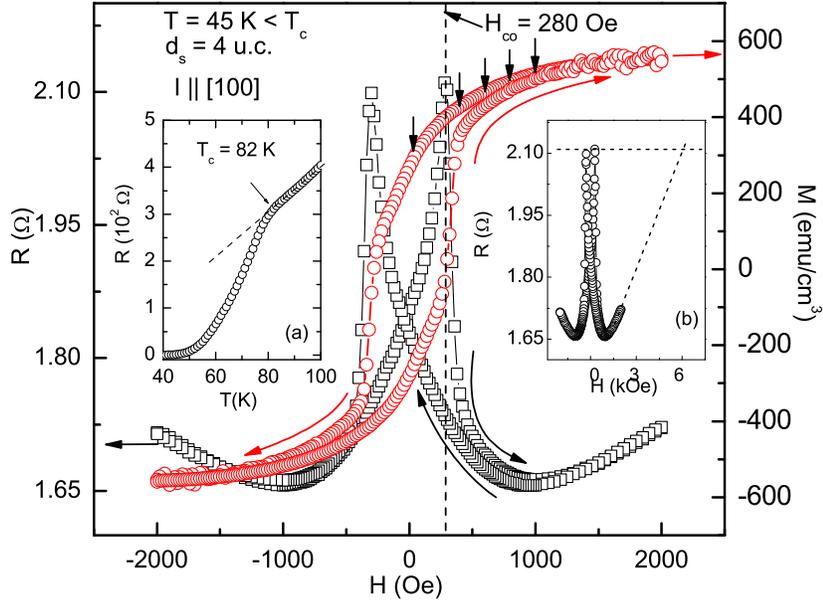}
\caption{\label{fig:epsart}  Applied magnetic Field $H$ dependent resistance $R$ (open squares) and magnetization
$M$ (open circles) of a LCMO/YBCO bilayer (the thickness $d_s$ of YBCO layer is 4 u.c.) measured in the mixed state of the bilayer, at a temperature $T$ of 45 K with $H$ along the [010] crystallographic direction. Insets: (a) $R-T$ curve measured in zero
field; (b) Same $R-H$ curve of the main panel measured over a wider $H$ range. }
\end{figure}

The  inset (b) to Fig. 1 is a plot of the same $R(H)$ data shown in the main panel, however, displayed over a larger field range. Notice that a linear extrapolation of the $R(H)$ data to high $H$ values shows that an in-plane  applied field of 6,000 Oe would give a resistance comparable to the peaks value. Therefore, a 6,000 Oe {\it in-plane} stray field in the domain walls is required to produce the measured peaks in $R(H)$.   However, the in-plane stray field  $H_{stray}^{ab}$ at the coercive field is much
less than the saturation magnetization ($M_{sat}^{ab}$) \cite{Thomas}. As a simple
estimate,  $H_{stray}^{ab} \approx 10\%\times 4\pi M_{sat}^{ab}=0.1\times 4\pi \times 566$ emu/cm$^3=710$ Oe. Hence, the $R(H)$ peaks are not due to an in-plane stray field since its estimated value of 710 Oe is much smaller than the required value of 6,000 Oe. Therefore, the $R(H)$ peaks can only be a result of an out-of-plane stray field. Such a conclusion is consistent with the sharp peaks in the $R(H)$ data since an out-of-plane stray field would give rise to a substantial number of vortices, hence, to sharp dissipation peaks in $R(H)$ due to vortex dissipation in the YBCO layer \cite{Bell}. 

These Bloch-type domain walls (the direction of the stray field that arises in the domain walls is out of plane) are a result of the cubic-to-tetragonal transition in  the SrTiO$_3$ substrate, which takes place below 105 K and induces twins in the LCMO layer  \cite{Vlasko-Vlasov}. The stresses in the twins  are the ones that induce the out-of-plane stray field \cite{Vlasko-Vlasov}. 

\subsection{Angular dependent magnetoresistance measurements}
Next, we  investigate how  $R(\varphi)$ evolves  when the magnetization changes
from the saturation state to the multi-domain state.  The arrows shown on the main plot of Fig. 1  mark the values of $H$ at which  $R(\varphi)$ was measured using the protocol corresponding to the upper curve of $M(H)$ of this figure.  The  in-plane angular dependent MR, defined as $R^{[100]}(\varphi)/R^{[100]}_{min}-1$,  is shown in Fig. 2(a). In these measurements, $I \parallel$ [100] crystallographic direction and  $R^{[100]}_{min}$ represents the minimum resistance. A fourfold symmetry is observed in the angular dependent MR data for $H$ rotated in the $ab-$plane of the bilayer at $T<T_c$. The  positions of the MR peaks are  slightly shifted from, but close to 90, 180, 270$^\circ$. Both this shift  and the magnitude of the MR peaks decrease with increasing $H$ from 400  to 1000 Oe.  

\begin{figure}
\centering
\includegraphics[width=0.5\textwidth]{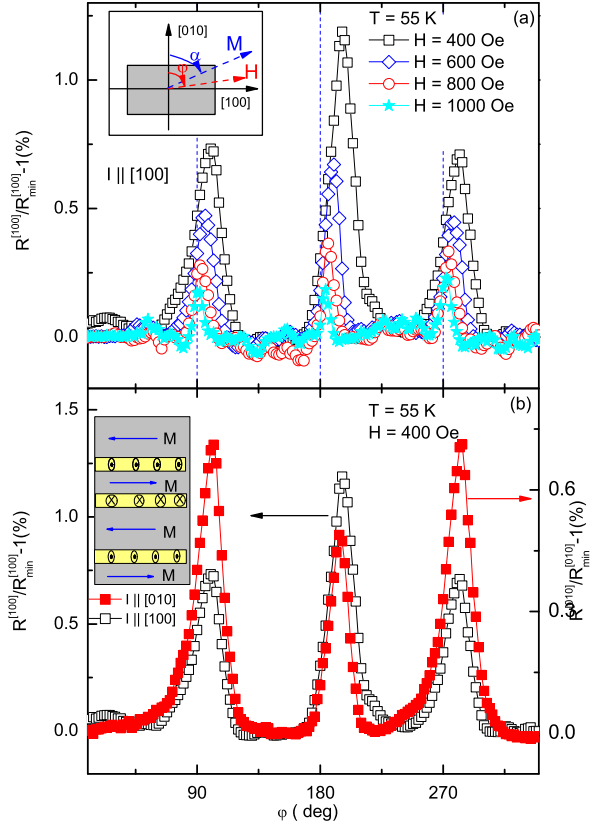}
\caption{\label{fig:epsart}  (a) Angular dependent magnetoresistance  $ R^{[100]}(\varphi)/R^{[100]}_{min}-1$ data  ($R_{min}^{[100]}$ is the minimum resistance) measured in the mixed state at 55 K and for applied magnetic fields of 400, 600, 800, and $1000$ Oe, with the current applied along the [100] crystallographic direction. Inset: top view of sample configuration. The magnetic field $H$ and magnetization $M$ are rotated in the ab-plane and make the angles $\varphi$ and $\alpha$, respectively, with the [010] crystallographic direction. 
 (b) Angular dependent  magnetoresistance  $ R^{[100]}(\varphi)/R^{[100]}_{min}-1$  and $ R^{[010]}(\varphi)/R^{[010]}_{min}-1$ data  measured in the mixed state at 55 K with the current applied along the [100] (open squares) and [010] (solid squares) crystallographic directions, respectively, and in an applied magnetic field of 400 Oe.  Inset:  top view of the stripe domain wall structure in the LCMO layer. The gray regions represent domains with the moments along the [100] and [\={1}00] directions and the yellow regions represent the domain walls with out-of-plane stray fields. }
\end{figure}

The equilibrium state of $M$ is achieved when the free energy $E$ of the system is minimum. Here, $E$ is the sum of Zeeman energy and magnetocrystal anisotropy energy (MAE) \cite{Riggs}; i.e.,
\begin{eqnarray}
E=-MHcos(\alpha-\varphi)+E_{MAE}(\alpha), 
\end{eqnarray}
where $\alpha$ is the
angle between $M$ and the [010] crystallographic direction [see inset to Fig. 2(a)]. If $H\ge H_{sat}$, the first term on the right-hand side of Eq.
(1) dominates, therefore $\alpha=\varphi$ gives the minimum $E$; hence $H$ and $M$ are along the same direction.   If $H \ll H_{sat,}$  the second term on the right-hand side of Eq. (1) dominates, therefore the minimum $E$ takes place
when $M$ is along the easy axis.  At intermediate $H$ values, both terms contribute to the energy $E$ and the relative angle between $H$ and $M$ directions for the equilibrium state is determined by the minimum value of E.  

The above discussion facilitates the understanding of the data of Fig. 2(a). Specifically, when $H=$ 1000 Oe,  the angular dependent MR data show maxima, corresponding to the maximum stray field for this applied magnetic field (maximum number of domain walls), at $\alpha \approx \varphi=90$, 180, or 270$^\circ$. The hard axes  for the LCMO/YBCO bilayer are the [010] and [100] crystallographic directions since the maximum number of domain walls takes place when the induced magnetization $M$ is along  the hard axis,   while the easy axes are in the diagonal directions. 

The small deviation of the MR peaks of Fig. 2(a) from the hard axes at lower values of $H$  is due to the fact that $M$ lags behind $H$ (the contribution of the magnetocrystal anisotropy energy can not be neglected), which is consistent with the fact that  this deviation becomes larger with decreasing $H$.
Also, the number of domains increases with decreasing $H$ from 1000 to 400 Oe, which produces an increase in the stray field with decreasing $H$.  As a result,  the value of the  MR peaks increases with decreasing $H$.

In addition, the magnitude of the MR peak depends on the angle between the $H$ and $I$ directions. Figure 2(b) gives the angular dependent MR for the current along [010] (solid symbols) and [100]
(open symbols) crystallographic directions.  Note that the
MR peak is always larger when $H\perp I $ ($\varphi$ is $90^{\circ}$ for the solid symbols and 180$^{\circ}$ for the open symbols) than when $H \parallel
I$ ($\varphi$ is 180$^{\circ}$ for the solid symbols and 90$^{\circ}$ for the open symbols). This change in the magnitude of the MR peaks with the angle between $H$ and $I$ reflects the interaction between  the stripe domain structure and   the vortex motion, as discussed below.  

The presence of stripe domains in the LCMO/YBCO bilayers has previously been reported  \cite{interface2, Laviano1, Laviano2}.  The inset to Fig. 2(b) is a sketch  of the cross section of the stripe domain structure in the LCMO layer at $T<T_c$ and at the coercive field. The gray regions represent the  stripe domains, and the yellow regions are the domain walls. The direction of the magnetization $M$ is shown along [100] and [\={1}00] and the directions of the stray field in the domain walls are also represented. Notice that adjacent domain walls have opposite directions of the stray field \cite{interface2, Laviano1,Laviano2}.  Hence, the out-of-plane stray field induces spontaneous vortices and anti-vortices in the YBCO layer \cite{interface2, Laviano1,Laviano2}. These flux  vortices  are driven by the Lorentz force and move in the direction  perpendicular to both $I$ and the stray field. Therefore, the smaller MR peak when
$I$ is along the stripes ($I$ $\parallel$ $H$) is due to the fact that the flux vortices are driven across the domain walls, which gives a smaller dissipation, hence, larger critical current of the superconducting film, due to the partial pinning of the vortices by the DWs. When $I$ is perpendicular to the stripes ($I\perp H$), the flux vortices are driven along the domain walls, thus their motion is not hindered by the DWs, hence the dissipation is larger and the critical current of the superconducting film smaller. 

This effect of the stripe domain walls on the critical current is based on the technique of pinning the flux vortices by the DWs rather than pinning the normal core of the vortices at the locally suppressed  superconductivity, realized by several possible  means (e.g., columnar defects, magnetic particles, etc.). The effect of the domain wall structure on the critical current had previously been studied in bilayers of a low $T_c$ superconductor (Nb) and an itinerant ferromagnet (SrRuO$_3$) [24]. Our  results are consistent with this study despite the different nature of the materials (e.g., $d$-wave vs. $s$-wave and half-metallic vs. itinerant ferromagnetic), suggesting that the interaction between the DWs and the flux vortices is independent of the nature of superconductivity and ferromagnetism. This is expected since the pinning of the flux vortices at the DWs is only a result of the magnetostatic interaction between the magnetic flux vortices and the magnetization of the FM layer.

The slight asymmetry at the base of the resistance peaks in Figs. 2(a) and 2(b) could be a result of the fact that $M$ lags behind $H$ when $H$ is rotated from an easy to a hard axis ($M$ prefers to lay along the easy axis), while $M$ jumps ahead of $H$, to the next
easy axis, when $H$ is rotated from a hard to an easy axis.

\section{Superconducting surface spin valve effect}
The magnetization $M_I$ of the LCMO layer within $2-3$ u.c. of the LCMO/YBCO interface is significantly different from the bulk magnetization, inside the LCMO layer \cite{interface1,interface2}.  In fact, polarized neutron reflectometry on YBCO/LCMO superlattices have shown strongly deppressed magnetization at the interface over 1 nm length scale \cite{Hoffmann}; i.e., the magnetic coupling near the LCMO/YBCO interface is very weak compared with the one of the bulk \cite{luo}, therefore the Curie temperature is expected to be less than the one of the bulk. So, the direction of $M_I$ could be tuned by the rotation of a small applied magnetic field, while not affecting the direction of the bulk magnetization. In this way, the parallel/antiparallel alignment of the surface and bulk magnetizations could be created. Therefore, this system is a good candidate for the investigation of the superconducting surface spin valve effect. 

 In order to investigate this effect in this F/S system, one, hence,  needs to pin the bulk magnetization of the LCMO layer and then use a low applied magnetic field that would control the magnetization at the LCMO/YBCO  interface.  Since  [110] is an easy axis, we applied $H$ in the [110] direction increasing its value up to $+2000$ Oe, to saturate the magnetization of  the LCMO layer along this direction, and then decreased $H$ to 35 Oe, so  the bulk magnetization remains pinned along this  [110] direction. We subsequently rotated the 35 Oe field in the $ab-$plane in order to rotate the magnetization $M_I$ of the surface layer, but not the bulk magnetization of the LCMO. [A small magnetic field cannot modulate the domain structure of the LCMO layer, hence the magnetocrystal anisotropy energy is the dominant term in Eq. (1). Nevertheless, it could rotates the magnetization $M_I$ of the LCMO/YBCO interface.]

Figure 3(a) shows the angular dependent MR data measured at 45 K $<T_c$ using the above protocol in both increasing (black square) and decreasing (red circle) angle. The fact that the angular dependent MR data are reversible shows that the bulk magnetization is pinned along the [110] easy axis, which gives the minimum energy of the system, without following the rotation of $H$.  The angular dependence of the resistance shown in Fig. 3(a)  is only observed in the superconducting state of the LCMO/YBCO bilayer. It could be a result of either domain nucleation at the surface layer and changes in their structure as a result of the motion of domain walls (a small field of 35 Oe could not have much effect on the domains of the bulk LCMO or on the superconductivity of the YBCO) or  of the rotation of the interface magnetization along with the 35 Oe field rotation, which would modulate the superconductivity at the LCMO/YBCO interface.  Below we show that the present data point toward the second  rather than the first scenario.

We measured also a minor magnetic loop using the following protocol. First, we applied the magnetic field along the [110] direction up to +2000 Oe, to again saturate the bulk LCMO magnetization along this direction. Then we decreased the field to zero and scanned $H$ over a small range; i.e. we increased the field to +40 Oe, then decreased it to $-40$ Oe, and then increased it back to zero. This obtained minor loop is shown in the inset to Fig. 3(b). The fact that this $M(H)$ minor loop is linear and reversible indicates magnetization rotation under the effect of an applied magnetic field (as opposed to domain nucleation, which would give rise to hysteresis). This is consistent with the proposed superconductive surface spin valve scenario and indicates that an exchange spring wall separates bulk and surface layers, as reported earlier  \cite{Prieto}. The existence of an exchange spring wall at the interface of the manganite is not surprising in view of a non homogeneous (depressed) magnetization  (see \cite{Hoffmann}). This layer
is typically a few nanometers thick, much thinner than the domain wall width, so that magnetization rotation within the thin layer is the most probable mechanism of magnetization reversal, a mechanism which  saves
exchange energy at the interface at the cost of Zeeman energy \cite{Prieto}.
 
\begin{figure}
\centering
\includegraphics[width=0.6\textwidth]{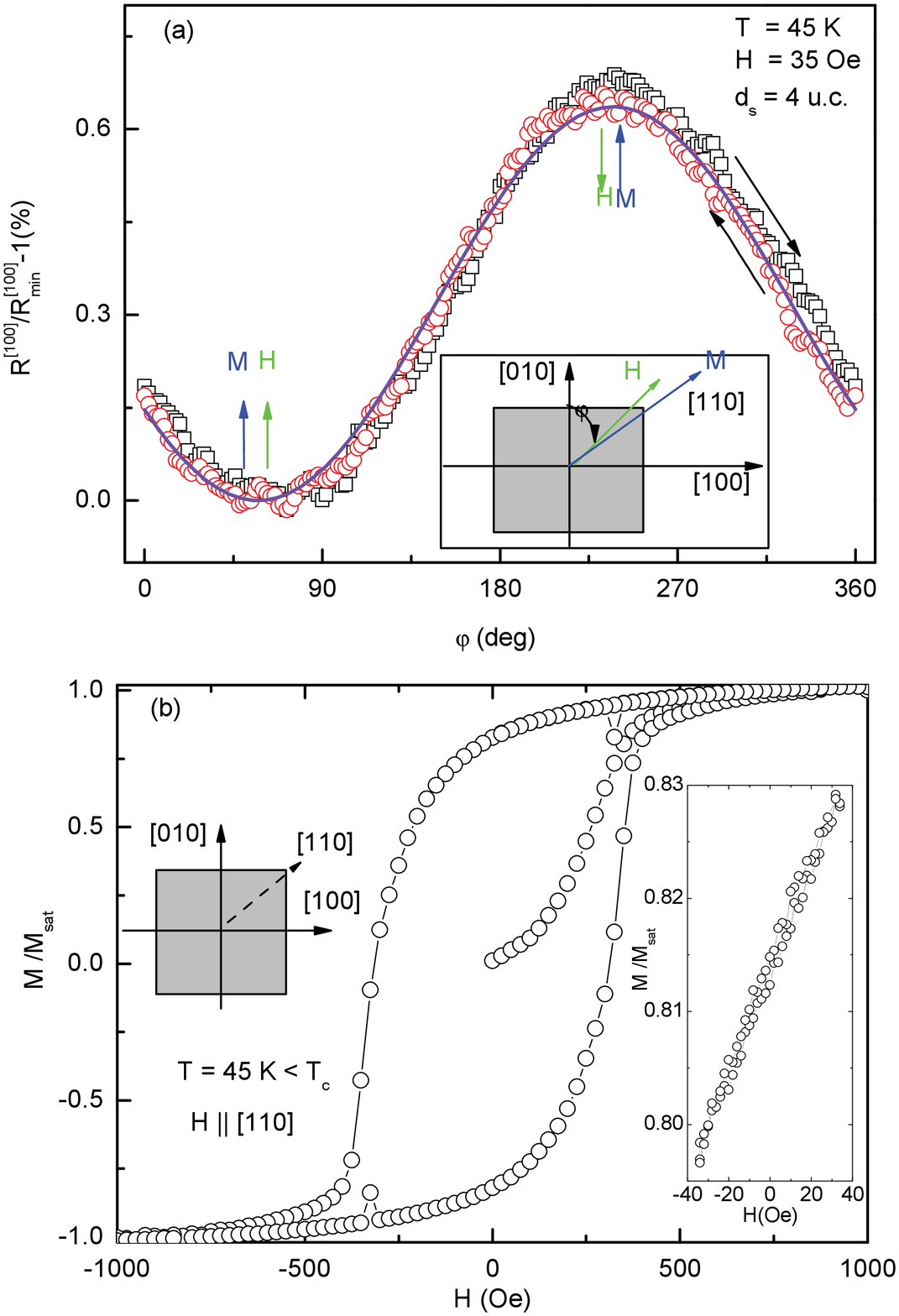}
\caption{\label{fig:epsart}   (a) Angular dependent magnetoresistance  $ R^{[100]}(\varphi)/R^{[100]}_{min}-1$ data  ($R_{min}^{[100]}$ is the minimum resistance) measured in the mixed state at 45 K and for an applied magnetic field $H$ of 35 Oe, with the current applied along the [100] crystallographic direction. Open circles and squares represent the data measured  clockwise and counterclockwise, respectively. Inset:  Sketch of the direction of bulk magnetization $M$ and of the applied magnetic field $H$. (b) Normalized magnetization $M/M_{sat}$ vs magnetic field $H$ applied along the [110] easy axis, measured in the mixed state at 45 K. Inset: Minor $M/M_{sat}$ vs $H$ loop measured after increasing $H$ up to 2000 Oe.}
\end{figure}

The MR curve of Fig. 3(a) is well fitted by $R^{[100]}(\varphi)/R_{min}^{[100]}-1=0.0067sin^{2}[(\varphi-57^0)/2]$
as shown by the solid curve in the figure. The fitting result of  $57^0$ suggests that the pinning angle of the bulk magnetization  is not exactly along the [110], but it makes an angle of $57^0$ with the [010] crystallographic direction. This $12^{\circ}$ difference between the [110] easy axis of LCMO and the pinning angle of the bulk magnetization could be due to  a small  tension or shape anisotropy of the bilayer in $a$ and $b$ directions. 

The superconductive magnetoresistive memory device has a structure similar to GMR memory devices. In fact, some of us have recently reported similar angular dependence of the magnetoresistance in F/S/F trilayers based on the same materials \cite{Visani}, but with magnetoreisistance values larger by more than one order of magnitude compared with the ones reported here. In Ref. \cite{Visani} it has been shown that this large magnetoresistance is tracking the relative alignment between top and bottom magnetic layers. On the other hand, the physics of superconducting memory devices, proposed based on the present data, is based on the S/F proximity effect \cite{Sangjun}; i.e., it is based on the oscillatory decay of the pair wave function predicted to occur in the ferromagnetic layer due to the influence of the exchange interaction on the Cooper pairs \cite{Buzdin}. Here we propose that the relative orientation between the surface and bulk magnetizations of the LCMO layer modulates the exchange interaction, hence, the spatial dependence of the Cooper pair wave function, therefore, the MR of the bilayer.  

A singlet Cooper pairs experiences less (more) pair breaking if the bulk magnetization of the LCMO layer and the magnetization of the LCMO/YBCO interface are antiparallel (parallel) since the different (same) sign of the exchange energies in the LCMO/YBCO interface and the LCMO layer makes the average of the exchange energy small (large) \cite{Sangjun}.
Nevertheless, the data of Fig. 3(a)  show that for $H || M$ ($\varphi=45^0$), MR
is minimum while for $H$ antiparallel to $M$ ($\varphi=225^0$), MR is maximum. Hence, these data suggest that triplet superconductivity is induced in the LCMO layer of the LCMO/YBCO system since the pair breaking effect is reduced when the  two ferromagnetic layers  are parallel \cite{Fominov}.
Furthermore, the Ginzburg Landau coherence length in the $c$ direction is about 0.2 nm at 45 K,  while the surface thickness is about 2-3 u.c. ($0.8-1.2$ nm) \cite{interface1,interface2,luo}. Therefore, the surface spin valve effect can not be due to singlet proximity effect, which is short range. Hence, one needs to consider a long range  proximity effect. 

The triplet created by a non-homogeneous magnetization at the interface of the S/F junction can produce  a long range  proximity effect \cite{tripletRMP,Volkov}. One possible source of inhomogeneity reported in the literature is domain walls. Volkov and Efetov have recently shown that starting from the {\it d}-wave superconductivity, the presence of domain walls perpendicular to the interface leads to the formation of both the singlet and odd triplet component of the  {\it s}-wave, and that the latter can penetrate the normal metal over long distances along the domain walls. However,  the  micronsize width of the domain walls in manganites is much larger that the nanometer scale coherence length. Also, the $M(H)$ loop shown in Fig. 3(b) measured along the [110] easy axis shows that domain walls are present in the bulk LCMO at zero field [the $M(H)$ loop is not really square], while the $M(H)$ minor loop of the inset to Fig. 3(b) shows no evidence of domains at the surface layer [the $M(H)$ loop is reversible].   So, this possible source of magnetic inhomogeneity is quite improbable.

Another source of non homogeneous (depressed) magnetization in this system could be phase segregation resulting from charge transfer or other interface related phenomena  \cite{Hoffmann}. Specifically, the depressed interfacial magnetization is (still)  laterally non uniform with a much shorter  nanometer length scale due to phase segregation. Inhomogeneities in the Mn$^{3+}$/Mn$^{4+}$ ratio resulting from charge transfer, strain relaxation, and  other interface processes that can not be microscopically followed by the La/Ca ratio are known to occur at manganite surfaces and interfaces. Charge spreads over the nanometer scale Thomas Fermi screening length to preserve charge neutrality, but nanometer scale phase separation occurs, giving rise to the stabilization of secondary phases and dead layers  with depressed magnetic and conducting properties \cite{Sun, Bibes, Giesen, Huijben}. Direct evidence of magnetic inhomogeneity has been recently found from magnetic force
microscopy \cite{Sun2}. This is most likely the source of the non homogeneous magnetization that gives rise to the triplet component and it is also the bases of the exchange spring surface layer, in which magnetization at the surface layer is weakly coupled to the bulk magnetization. 

It has been proposed theoretically that unpolarized supercurrents could be converted to  triplet-pairing at spin-active interfaces \cite{Eschrig1, Eschrig2}, while it has been found experimentally that there is a 100 nm thick layer at the LCMO/YBCO interface that displays a suppressed (but non-zero) ferromagnetic moment  \cite{Chakhalian}. Therefore, this could also be a possible source  for the triplet component present at the LCMO/YBCO interface.

Regardless its origin, our finding of triplet superconductivity is consistent  with recent reports of triplet superconductivity in LCMO/YBCO/LCMO trilayers \cite{Hu} and bilayers \cite{Kalcheim}. This triplet condensate would gives rise to the observed one-fold symmetry: maximum resistance when moments are antiparallel and minimum resistance when moments are parallel. These results of triplet superconductivity in a ferromagnetic manganite and unconventional superconductor complement the results on spin-triplet superconductivity found at interfaces  of hybrids of ferromagnets (such as CrO$_2$, Ho and Co) and conventional superconductors \cite{Keizer, Sosnin, Khaire, Robinson}.

\section{Conclusions} 
We performed  magnetoresistance MR and magnetization measurements on LCMO/YBCO bilayers below the superconducting transition temperature $T_c$ of the bilayers and studied in detail their spin-dependent transport. We showed that the vortex dissipation, related with the out-of-plane stray field induced by the stresses in the twins of the LCMO layer as a result of a structural phase transition in the substrate, gives MR peaks at the coercive field. More interestingly,  at low magnetic field values, we found a novel superconducting surface spin valve effect, in which the MR signal depends on the relative angle between the magnetizations of the LCMO/YBCO interface and the LCMO layer;  i.e. MR is minimum when the two magnetizations are parallel and maximum when they are antiparallel.  Our study on the spin dependent transport in ferromagnetic/superconductor bilayers opens a new avenue to the realization of spintronic devices. This novel superconducting surface spin valve effect can be used in a superconductive magnetoresistive memory device as a magnetoresistive switching element.

\ack

This research was supported by the National Science Foundation under Grant No. DMR-1006606 at KSU and MCYT MAT 2008-06517 at U. Complutense de Madrid. T. H. acknowledge the support of ICAM Branches Cost Sharing Fund from the Institute for Complex
Adaptive Matter.


\section*{References}
\begin{thebibliography}{35}
\bibitem{Prinz}
Prinz Gary A 1998 
Magnetoelectronics
 {\it Science}  {\bf 282} 1660

\bibitem{Baibich}
 Baibich M N,  Broto J M, Fert A, Nguyen Van Dau F, Petroff F, Etienne P,
  Creuzet G, Friederich A and Chazelas J 1998
 Giant magnetoresistance of (001)Fe/(001)Cr magnetic superlattices
{\it Phys. Rev. Lett.} {\bf 61} 2472

\bibitem{Tagirov}
Tagirov L R 1999
Low-field superconducting spin switch based on a superconductor/ferromagnet multilayer
 {\it Phys. Rev. Lett.} {\bf 83} 2058

\bibitem{Gu}
Gu J Y, You C -Y, Jiang J S, Pearson J, Bazaliy Ya B and Bader S D  2002
Magnetization-orientation dependence of the superconducting transition temperature in the ferromagnet-superconductor-ferromagnet system: CuNi/Nb/CuNi
{\it Phys. Rev. Lett.} {\bf 89} 267001

\bibitem{Potenza}
Potenza A and Marrows C H 2005
Superconductor-ferromagnet CuNi/Nb/CuNi trilayers as superconducting spin-valve core structures
{\it Phys. Rev. B} {\bf 71} 180503(R)

\bibitem{Moraru}
Moraru Ion C, Pratt W P, Jr., and Birge Norman O 2006
Magnetization-dependent $T_c$ shift in ferromagnet/superconductor/ferromagnet trilayers with a strong ferromagnet
{\it Phys. Rev. Lett.} {\bf 96} 037004 


\bibitem{Yanson}
Yanson I K,  Naidyuk Yu G, Fisun  V V, Konovalenko A, Balkashin O P, Triputen Yu L and Korenivski V 2007
Surface spin-valve effect
{\it Nano Lett.} {\bf 7} 927

\bibitem{interface1}
Stahn J, Chakhalian J, Niedermayer C, Hoppler J, Gutberlet T, Voigt J,
  Treubel F,  Habermeier H-U, Cristiani G, Keimer B and Bernhard C 2005
 Magnetic proximity effect in perovskite superconductor/ferromagnet
  multilayers
{\it Phys. Rev. B} {\bf 71} 140509(R)

\bibitem{interface2}
Chakhalian J, Freeland J W, Srajer G, Strempfer J, Khaliullin G, 
  Cezar J C, Charlton T, Dalgliesh R, Bernhard C, Cristiani G, 
  Habermeier H-U and Keimer B 2006
 Magnetism at the interface between ferromagnetic and superconducting
  oxides
{\it Nature Phys.} {\bf 2} 244

\bibitem{interface3}
 Santamaria J 2006
Complex oxides: Interfaces on stage
{\it Nature Phys.} {\bf 2} 229

\bibitem{Steiner}
Steiner R and Ziemann P
 Magnetic switching of the superconducting transition temperature in
  layered ferromagnetic/superconducting hybrids: Spin switch versus stray field
  effects
 {\it Phys. Rev. B} {\bf 74} 094504

\bibitem{Rusanov}
Rusanov A Y, Hesselberth M, Aarts J and Buzdin A I 2004
Enhancement of the superconducting transition temperature in
  Nb/permalloy bilayers by controlling the domain state of the ferromagnet
{\it Phys. Rev. Lett.} {\bf 93} 057002

\bibitem{Rusanov2}
Rusanov A Y, Habraken S and Aarts J 2006
Reverse spin switch effects in ferromagnet-superconductor-ferromagnet
  trilayers with strong ferromagnets
{\it Phys. Rev. B} {\bf 73} 060505(R)

\bibitem{Buzdin}
Buzdin A I 2005
Proximity effects in superconductor-ferromagnet heterostructures
{\it Rev. Mod. Phys.} {\bf 77} 935

\bibitem{Bell}
Bell C, Tursucu S and Aarts J 2005
Flux-flow-induced giant magnetoresistance in all-amorphous
  superconductor-ferromagnet hybrids
{\it Phys. Rev. B}, {\bf 74} 214520

\bibitem{Pena2006}
Pe$\tilde{n}$a V, Visani C, Garcia-Barriocanal J, Arias D, Sefrioui Z,
  Leon C, and Santamaria J 2006
Spin diffusion versus proximity effect at ferrromagnet/superconductor
  $la_{0.7}ca_{0.3}mno_3$/$yba_2cu_3o_{7-\delta}$ interafecs
{\it Phys. Rev. B} {\bf 73} 104513

\bibitem{Pena2004}
Pe$\tilde{n}$a V, Sefrioui Z, Arias D, Leon C, Santamaria J, Varela M,
  Pennycook S J and Martinez J L 2004
 Coupling of superconductors through a half-metallic ferromagnet:
  Evidence for a long-range proximity effect.
{\it Phys. Rev. B} {\bf 69} 224502

\bibitem{xiaoPRB}
Xiao H, Hu T, Almasan C C, Sayles T A and Maple M B 2008
{\it Phys. Rev. B} {\bf 78} 014510

\bibitem{Thomas}
Thomas L, Samant M G and Parkin S S P 2000
 Domain-wall induced coupling between ferromagnetic layers
{\it Phys. Rev. Lett.}  {\bf 84} 1816

\bibitem{Vlasko-Vlasov}
Vlasko-Vlasov V K, Lin Y K, Miller D J, Welp U, Crabtree G W and
  Nikitenko V I 2000
Direct magneto-optical observation of a structural phase transition
  in thin films of manganites
{\it Phys. Rev. Lett.} {\bf 84} 2239

\bibitem{Riggs}
Riggs K T, Dahlberg E D, and Prinz G A 1990
First-order magnetic-field-induced phase transition in epitaxial iron
  films studied by magnetoresistance
{\it Phys. Rev. B} {\bf 41} 7088

\bibitem{Laviano1}
Laviano F, Gozzelino L, Mezzetti E, Przyslupski P, Tsarev A and
  Wisniewski A 2005
Control of the vortex movement and arrangement by out-of-plane
  magnetic structures in twinned YBa$_2$Cu$_3$O$_{7-x}$/La$_{0.67}$Sr$_{0.33}$MnO$_3$ bilayer
{\it Appl. Phys. Lett.} {\bf 86} 152501

\bibitem{Laviano2}
Laviano F, Gozzelino L,  Gerbaldo R, Ghigo G, Mezzetti E, 
  Przyslupski P, Tsarou A and Wisniewski A 2007
Interaction between vortices and ferromagnetic microstructures in
  twinned cuprate/manganite bilayers
{\it Phys. Rev. B} {\bf 76} 214501

\bibitem{Feigenson}
Feigenson M, Klein L, Karpovski M, Reiner J W and Beasley M R 2005
Suppression of the superconducting critical current of Nb in bilayers of Nb/SrRuO$_3$
{\it J. Appl. Phys.} {\bf 97} 10J120

\bibitem{Hoffmann}
Hoffmann A, teVelthuis S G E, Sefrioui Z, Santamaria J, 
Fitzsimmons M R, Park S and Varela M 2005
Suppressed magnetization in La$_{0.7}$Ca$_{0.3}$MnO$_3$/YBa$_2$Cu$_3$O$_{7-\delta}$ superlattices
{\it Phys. Rev. B} {\bf 72} 140407(R)

\bibitem{luo}
Luo W, Pennycook S J and Pantelides S T 2008
Magnetic "Dead" layer at a complex oxide interface
{\it Phys. Rev. Lett.} {\bf 101} 247204

\bibitem{Prieto}
 Prieto Jose L, van Aken Bas B, Martin Jose I, Perez-Junquera A, 
  Burnell Gavin, Mathur Neil D and Blamire Mark G 2005
Absence of spin scattering of in-plane spring domain walls
{\it Phys. Rev. B} {\bf 71} 214428

\bibitem{Visani}
Visani C, Nemes N M, Rocci M, Sefrioui Z, Sefrioui C, Leon C, 
  te Velthuis S G E, Hoffmann A and Fitzsimmons M R 2000
{\it Phys. Rev. B} {\bf 81} 094512

\bibitem{Sangjun}
Oh S, Youm D and Beasley M R 1997
A superconductive magnetoresistive memory element using controlled
  exchange interaction
{\it Appl. Phys. Lett} {\bf 71} 2376

\bibitem{Fominov}
 Fominov Y V, Golubov A A and M Y Kupriyanov
Triplet proximity effect in fsf trilayers 2003
{\it JETP Letters} {\bf 77} 510

\bibitem{tripletRMP}
Bergeret F S,  Volkov A F, and Efetov K B 2005
Odd triplet superconductivity and related phenomena in
  superconductor-ferromagnet structures
{\it Rev. Mod. Phys.} {\bf 77} 1321

\bibitem{Volkov}
Volkov A F and Efetov K B 2009
Proximity effect and its enhancement by ferromagnetism in high-temperature superconductor-ferromagnet structures
{\it Phys. Rev. Lett.} {\bf 102} 077002

\bibitem{Sun}
Sun J Z, Abraham D W, Rao R A and Eom C B 1999
Thickness-dependent magnetotransport in ultrathin manganite films
{\it Appl. Phys. Lett.} {\bf 74} 3017

\bibitem{Bibes}
Bibes M, Balcells L, Valencia S, Fontcuberta J, Wojcik M, Jedryka E and
  Nadolski S 2001
Nanoscale multiphase separation at La$_{2/3}$Ca$_{1/3}$MnO$_3$/SrTiO$_{3}$ interfaces
{\it Phys. Rev. Lett.} {\bf 87} 067210

\bibitem{Giesen}
Giesen F, Damaschke B, Moshnyaga V, Samwer K and Muller G A 2004
Suppression of interface-induced electronic phase separation in all-manganite multilayers by preservation of the Mn-O chain network
{\it Phys. Rev. B} {\bf 69} 014421

\bibitem{Huijben}
Huijben M, Martin L W, Chu Y H, Holcomb M B, Yu P, Rijnders G,
 Blank D H A and Ramesh R 2008
Critical thickness and orbital ordering in ultrathin La$_{0.7}$Sr$_{0.3}$MnO$_3$ films
{\it Phys. Rev. B} {\bf 78} 094413

\bibitem{Sun2}
Sun Y H, Zhao Y G, Tian H F, Xiong C M, Xie B T, Zhu M H, Park S,
  Wu W, Li J Q and Li Qi 2008
Electric and magnetic modulation of fully strained dead layers in La$_{0.67}$Sr$_{0.33}$MnO$_{3}$ films
{\it Phys. Rev. B} {\bf 78} 024412

\bibitem{Eschrig1}
M. Eschrig, J. Kopu, J. C. Cuevas, and Gerd Schon 2003
Theory of half-metal/superconductor heterostructures
{\it Phys. Rev. Lett.} {\bf 90} 137003 

\bibitem{Eschrig2}
Eschrig Matthias and Lofwander Tomas 2008
Triplet supercurrents in clean and disordered half-metallic ferromagnets
{\it Nature Phys.} {\bf 4} 138

\bibitem{Chakhalian}
J. Chakhalian and J. W. Freeland and G. Srajer and J. Strempfer and G. Khaliullin and J. C. Cezar and T. Charlton and R. Dalgliesh and C. Bernhard and G. Cristiani and H. -U. Habermeier and B. Keimer 2006
Magnetism at the interface between ferromagnetic and superconducting oxides
{\it Nature Phys.} {\bf 2} 244

\bibitem{Hu}
Hu T, Xiao H, Visani C, Sefrioui Z, Santamaria J and Almasan C C 2009
Evidence from magnetoresistance measurements for an induced triplet
  superconducting state in
  la$_{0.7}$Ca$_{0.3}$MnO$_3$/YBa$_2$Cu$_3$O$_{7-\delta}$ 
{\it Phys. Rev. B} {\bf 80} 060506(R)

\bibitem{Kalcheim}
Kalcheim Yoav, Kirzhner Tal,  Koren Gad and Millo Oded 2010
Long range proximity effect in La$_{2/3}$Ca$_{1/3}$MnO$_3$ (LCMO)/(100)YBa$_2$Cu$_3$O$_{7-\delta}$(YBCO) ferromagnet/superconductor bilayers: Evidence for induced triplet superconductivity in the ferromagnet
{\it cond-mat.supr-con arXiv:1010.0390v1} {\bf 78} 024412

\bibitem{Keizer}
Keizer R S, Goennenwein S T B, Klapwijk T M, Miao G, Xiao G and Gupta A 2006
A spin triplet supercurrent through the half-metallic ferromagnet CrO$-2$
{\it Nature} {\bf 439} 825

\bibitem{Sosnin}
Sosnin I, Cho H and Petrashov V T and Volkov A F 2006
Superconducting phase coherent electron transport in proximity conical ferromagnets
{\it Phys. Rev. Lett.} {\bf 96} 157002

\bibitem{Khaire}
Khaire Trupti S, Khasawneh Mazin A, Pratt W P, Jr., and Birge Norman O 2010
Observation of spin-triplet superconductivity in Co-based Josephson junctions
{\it Phys. Rev. Lett.} {\bf 104} 137002

\bibitem{Robinson}
Robinson J W A, Witt J D S, Blamire M G 2010
Controlled injection of spin-triplet supercurrents into a strong ferromagnet
{\it Science} {\bf 329} 59



\end{thebibliography}
\end{document}